\documentclass{jpsj2}
\title{The 1D Bose Gas with Weakly Repulsive Delta Interaction}

\author{\textsc{M.T. Batchelor}$^{1}$,
 \textsc{X.W. Guan }$^{1}$, \textsc{C. Dunning}$^{2}$ and  \textsc{J. Links}$^{2}$
}

\inst{$^{1}$    Department of Theoretical Physics, RSPSE, and\\
Mathematical Sciences Institute\\
The Australian National University, Canberra ACT 0200, Australia\\
$^{2}$ Centre for Mathematical Physics, School of Physical Sciences,\\
The University of Queensland, Brisbane 4072, Australia}

\abst{
We consider the asymptotic solutions to the Bethe ansatz equations of the
integrable model of interacting bosons in the weakly interacting limit.
In this limit we 
establish that the ground state maps to the highest energy 
state of a strongly-coupled repulsive bosonic pairing model.  
}

\kword{Integrable models, 1D Bose gas, Bethe ansatz equations, BCS pairing models}

\def\e{\epsilon}

\def\s{\sigma}

\def\beq{\begin{equation}}
\def\eeq{\end{equation}}
\def\bea{\begin{eqnarray}}
\def\eea{\end{eqnarray}}
\def\ba{\begin{eqnarray}}
\def\ea{\end{eqnarray}}

\def\nn{\nonumber}

\def\S{\mathcal{S}}
\def\L{\mathcal{L}}

\begin{document}
\maketitle

\section{Introduction} 

There has been a revival of interest in the exactly solved 1D model of
interacting bosons \cite{LL}.
In part this is due to the experimental realisation of a
quasi-1D quantum gas of bosons at ultracold 
temperatures.\cite{EXP1,EXP2,EXP3,EXP4,EXP5}
The essential point is that the interactions between the trapped atoms
can be tuned to bring about a continuous passage from the weakly interacting regime
to the strongly interacting regime.
In this way the full subtleties of quantum many-body physics are observed, from 
Bose-Einstein condensation in the weak coupling regime to the
pronounced fermionic behaviour of the Tonks-Girardeau gas in  
the strong coupling regime.

Recently it was observed \cite{bgm} with the help of numerical
analysis that the Bethe ansatz roots for the ground state of the
exactly solved 1D Bose gas in the weak coupling limit satisfy 
a similar set of equations as an exactly solved BCS boson pairing model
in the strong coupling limit.
Here we further clarify this correspondence in two ways:
first we derive, for arbitrary particle number, the system of
equations satisfied by the ground state roots of the 1D Bose gas in the
weak coupling limit.  
Then we show that the mapping to a BCS type
system is precise for certain bosonic systems where the Cooper pairs
are formed from integer spin particles.

\section{Weak coupling limit}

The Hamiltonian 
\bea &&{\cal H} =-\sum_{j=1}^N \frac{\partial^2}{\partial x_j^2}
+2c\sum_{j<k}^N \delta(x_j-x_k)   
\eea
describing $N$ interacting bosons on a periodic interval of length $L$ has been
extensively studied  \cite{LL,L, MC, YY, Gaudin, MAT, KOR}.
The eigenstates have energy and momenta  given by
\beq
{\cal E}=\sum_{j=1}^N k_j^2, \qquad 
P=\sum_{j=1}^N k_j,
\eeq
where $k_j$ satisfy the Bethe ansatz equations (BAE) 
\beq
\exp( \mathrm{i} k_jL)=-\prod_{l=1}^N\frac{k_j-k_l+\mathrm{i}c}
{k_j-k_l-\mathrm{i}c}, \qquad j=1,\ldots,N. \label{BA}  
\eeq   
For repulsive interactions ($c>0$) it is known that the Bethe roots $k_j$ 
are real  and distinct \cite{KOR}. 
Moreover, the eigenspectrum is positive, i.e., 
${{\cal E}_0} \le {\cal E}_1 \le {\cal E}_2 \dots$, where ${\cal E}_0$ is the
ground state.

Asymptotic solutions to the BAE (\ref{BA}) are possible in two limiting cases: 
$Lc\ll1$ and $Lc\gg1$, corresponding to weak and strong delta interaction.  
In the limit $Lc\gg1$ \cite{bgm}, which we do not discuss here, 
the asymptotic solutions to the BAE (\ref{BA}) describe the Tonks-Giradeau gas \cite{Tonks,KG}.
In the limit $Lc\ll1$, numerical checks suggest that the momenta $k_j$ are proportional to the square root of $c$ \cite{bgm}.
It follows that,  to order $c^2$, the BAE (\ref{BA}) reduce to 
\begin{eqnarray}
\exp( \mathrm{i} k_jL)&\approx&1-2\sum^{N}_{\ell=1}\frac{c^2}{(k_j-k_{\ell})^2}-4\sum_{\ell=1}^{N-1}\sum_{\ell < \ell '=2}^{N}\frac{c}{(k_j-k_{\ell})}\frac{c}{(k_j-k_{\ell '})}
\nonumber\\
& &
+ \,\, \mathrm{i} \sum_{\ell=1}^N\frac{2c}{(k_j-k_{\ell})},
\end{eqnarray}
in which the summations exclude $\ell=j$ and $\ell'=j$. 
Indeed, to order $c^2$,
\begin{eqnarray}
\cos(k_j L)&\approx & 1-2\sum^{N}_{\ell=1}\frac{c^2}{(k_j-k_{\ell})^2}-4\sum_{\ell=1}^N\sum_{\ell < \ell '=2}^{N}\frac{c}{(k_j-k_{\ell})}\frac{c}
{(k_j-k_{\ell '})},\label{BE-app1}\\
\sin(k_j L)& \approx & \sum_{\ell=1}^N\frac{2c}{(k_j-k_{\ell})}.
 \label{BE-app2}
\end{eqnarray}
The solution of which determines the asymptotic roots of the BAE (\ref{BA}).
>From Eqs. (\ref{BE-app1}) and (\ref{BE-app2}) the Bethe roots are seen to satisfy 
\begin{equation}
k_j=\frac{2\pi d_j}{L}+\frac{2c}{L} \, {\sum_{\ell\ne j}^N} \frac{1}{k_j-k_\ell}, 
\qquad j=1,\ldots N.
\label{momenta}
\end{equation}
Here $d_j=0,\pm 1,\pm 2, \ldots$
denotes excited states and the summation excludes $j=\ell$. 

The asymptotic equations are closely related to those appearing in 
Stieltjes problems \cite{SD}. 
The ground state has zero total momentum with $d_j=0$ for $j=1,\ldots,N$, with
the ground state energy per particle 
\begin{equation}
\frac{{\cal E}_0}{N}=\frac{c(N-1)}{L},
\label{enrgy1}
\end{equation}
following directly from Eq.  (\ref{momenta}).
The algebraic equations (\ref{momenta}) for the ground state
are given in Gaudin \cite{Gaudin}.
In this way Gaudin showed that the $k_j$ are roots
of Hermite polynomials of degree $N$, namely $H_N(k)=0$.
They are also related to roots of the Laguerre polynomial.\cite{bgm}
These connections provide a systematic way for studying
quantities such as the momentum distribution function and correlations.
The normalised momentum density distribution is given by the semi-circle law 
\cite{Gaudin}
\begin{equation}
n(k)=\frac{L}{2N\pi c}\left(4c\rho-k^2\right)^{\frac{1}{2}}, 
\label{density}
\end{equation}
where $\rho=N/L$.
The stronger the interaction strength, the larger the momentum distribution region. 
This reveals a significant signature of the 1D Bose gas in the weakly repulsive regime $Lc\ll1$. 
Remarkably, this behaviour was recently observed in the
experiments for weakly interacting bosons.\cite{EXP1,EXP2,EXP3,EXP4,EXP5}
If $c=0$ all the particles condense in the ground state at zero temperature.

Now consider the excitations above the ground state, which have 
total momentum $P={2n\pi}/{L}$ with $n=0,\pm 1,\pm 2, \ldots$.
As an example of the solutions obtained from the asymptotic equations 
(\ref{momenta}), consider the numerical data for four bosons given in  {Table I}.
Each state is characterised by the quantum numbers $d_j$, with
total momentum $P=\sum_j^N {2d_j\pi}/{L}$.
The excitation energies are at least doubly degenerate in view of the
counterpart assignments $- d_j$ with momenta $- k_j$.
Shown for direct comparison are the numerical solutions from the full BAE (\ref{BA}).
Clearly the agreement is excellent.

Let us consider the lowest excited state ${\cal E}_1$, with total 
momentum $P={2\pi }/{L}$, in more detail. 
Without loss of generality,  we choose an assignment $d_1=1$ and $ d_j=0$ for $j=2,\ldots, N$. 
Using Eq.~(\ref{momenta}) we can approximately calculate the lowest excitation energy.
Specifically, 
\begin{equation}
\frac{{\cal E}_1}{N} =\sum_{j=1}^Nk_j^2
\,\, \approx  \,\, \frac{2\pi k_1}{LN} + \frac{c(N-1)}{L}, 
\end{equation} 
from which we can infer that  
\begin{eqnarray}
k_1& \approx & \frac{2\pi}{L}+\frac{c(N-1)}{\pi},\label{k1} \\
\frac{{\cal E}_1}{N} &\approx & \frac{c(N-1)}{L}+\frac{4\pi^2}{L^2N}+\frac{2c(N-1)}{LN}, 
\label{E1}
\end{eqnarray}
which are valid for $N/L$ finite.
These approximations are superior to those given earlier \cite{bgm}.
It is clearly seen from Eq.~(\ref{E1}) that the energy gap will vanish in the limit $N \to \infty$. 
For the values of Table I, 
Eqs.~(\ref{k1}) and (\ref{E1}) give $k_1\approx 6.30705$ and 
${{\cal E}_1} \approx 39.9284$, which are
in good agreement with the numerical results obtained from 
Eqs.~(\ref{momenta}) and (\ref{BA}). 
As a further example, consider $N=8$ with $c=0.025$ and $L=2$.
In this case Eqs.~(\ref{k1}) and (\ref{E1}) give  $k_1\approx 3.1973$
and ${{\cal E}_1} \approx 10.7446$.
These results are to be compared with $k_1\approx 3.1693$ and 
${{\cal E}_1} \approx 10.3065$,
which follow from Eq.~(\ref{momenta}), and  $k_1\approx 3.1966$ and
${{\cal E}_1} \approx 10.7403$, obtained from the BAE (\ref{BA}).

\begin{center}
\begin{table}
\caption{The ground state and the leading five excitations of four weakly
interacting bosons with $L=1$ and $c=0.025$. 
For each state, solutions of the asymptotic equations (\ref{momenta}) 
are shown on the first line.
The second line shows solutions obtained from the full BAE (\ref{BA}).
}
\vskip 3mm
\begin{tabular}{|c|c|c|c|c|c|c|c|}
\hline 
$n$&${\cal E}_n$&$P$&$k_4$&$k_3$&$k_2$&$k_1$&$(d_1,d_2,d_3,d_4)$\\
\hline 
0&
0.30000&
0&
-0.36910&
-0.11731&
 0.11731& 
0.36910 &
(0,0,0,0)
\\
0&
0.29938&
0&
-0.36872&
-0.11719&
0.11719& 
0.36872 &
\\
\hline
1&
39.88783&
2$\pi $&
-0.28162&
-0.00793& 
0.26576&
6.30379&
(1,0,0,0)
\\
1&
39.92754&
2$\pi $&
-0.28133&
-0.00792&
 0.26548& 
6.30697&
\\
\hline
2&
79.37604&
4$\pi $&
-0.17378&
0.14205&
6.13794&
 6.45378& 
(1,1,0,0)
\\
2&
79.45587&
4$\pi $&
-0.17361&
0.14190&
6.14129&
6.45679& 
\\
\hline
3&
79.42614&
0 &
-6.29985&
-0.15791&
0.15791&
 6.29985& 
(1,-1,0,0)
\\
3&
79.50607&
0 &
-6.30303&
-0.15775&
0.15775&
6.30303& 
\\
\hline
4&
118.7646&
6$\pi $&
-0.02380&
6.01424&
6.28793&
 6.56162& 
(1,1,1,0)
\\
4&
118.8844&
6$\pi $&
-0.02378&
6.01771&
6.29111&
6.56452& 
\\
\hline
5&
118.8647&
2$\pi$ &
-6.29590&
-0.00793&
6.13393&
6.44990& 
(1,1,-1,0)
\\
5&
118.9846&
2$\pi$ &
-6.29907&
-0.00793&
6.13727&
6.45292& 
\\
\hline 
\end{tabular}
\end{table}
\end{center}

\section{Link to the BCS model}

The weak coupling limit provides a direct link between the integrable 1D Bose gas and 
the integrable BCS pairing models.
In this regime the Bethe roots
for the ground state are of the form
%
\begin{equation}
k_1 = -k_2 =\sqrt{E_1},\,k_3=-k_4=\sqrt{E_2},\ldots, \, 
k_{2M-1}=-k_{2M}=\sqrt{E_M},\label{P3}
\end{equation}
where the $E_i$ satisfy the equations (cf (\ref{momenta}))
\begin{equation}
-\frac{L}{2c} + {\sum_{j\ne i}^{M}} \frac{2}{E_i-E_j}=-\frac{1}{2E_i},
\label{bose}
\end{equation}
for $ i = 1, \ldots, M$ and  the total number of bosons are even, i.e. $N=2M$. 

For an odd number of  bosons, $N=2M+1$, the ground state Bethe roots are given by
%
\begin{equation}
k_1=0,\,k_{2}=-k_{3}=\sqrt{E_1},\,k_{4}=-k_{5}=\sqrt{E_2},
\ldots, k_{2M}=-k_{2M+1}=\sqrt{E_M},\label{P4}
\end{equation}
where the ${E_i}$ satisfy the equations
\begin{equation}
{\sum_{j\ne i}^{M}} \frac{2}{E_i-E_j}-\frac{L}{2c}=-\frac{3}{2E_i},
\label{bose2}
\end{equation}
for $ i = 1, \ldots, M$.

Similar equations have arisen in a number of contexts \cite{BCS1,yba,BCS3,BCS5,SD}.
Of particular interest here is the connection between Eq. (\ref{bose})
and Richardson's equations for the BCS pairing model \cite{BCS3,BCS4}
in the strong coupling limit \cite{yba}.  
We now make the connection with these `pairing interaction' Hamiltonians more precise. 
In terms of the $su(2)$ algebra with commutation relations 
\beq [S^z, S^{\pm}] = \pm S^\pm, \qquad  [S^+,S^-] =2 S^z, \label{su2} 
\eeq
we consider the class
of Hamiltonians acting on the $\L$-fold tensor product of
$su(2)$-modules, (not necessarily finite-dimensional,) with lowest
weight $-\S$.
Specifically,
\bea 
{\cal H}&=& \sum_{j=1}^{\L}2\e_j(S_j^z+\S
I)-g\sum_{j,k=1}^{\L}S_j^+S_k^-,
\label{ham} 
\eea
where $I$ is the identity operator.
The single particle energy levels $\e_j$ and the coupling constant $g$ are arbitrary real  
parameters.
The energy levels of the
Hamiltonians (\ref{ham}) are ${\cal E}=2\sum_{j=1}^m E_j$ where the
$\{E_j\}$ satisfy the Bethe equations \cite{lzmg}
\beq
\frac{2}{g}+ \sum_{k=1}^{\L}\frac{2\S}{E_j-\e_k}
=\sum_{\ell\neq j}^m \frac{2}{E_j-E_\ell}.
\label{bcsbae} 
\eeq

Suppose each energy level $\e_j$ can be occupied by a
spin $s$ particle so the degeneracy is $2s+1$. We introduce the creation and
annihilation operators $a_{\sigma},\,a_{\sigma}^\dagger$, where
$\sigma=-s,-s+1,...,s$ is the spin label, satisfying the usual
boson commutation (resp. fermion anticommutation) relations for integer (resp.
half-odd integer) values of $s$. 
We can define a representation
of $su(2)$ through
\bea && 
S^-=-\frac{1}{2}\sum_{\sigma=-s}^s 
(-1)^{s-\sigma}a_{\sigma}a_{\overline{\sigma}},
~~~
S^+=\frac{1}{2}\sum_{\sigma=-s}^s (-1)^{s-\sigma} a_{{\sigma}}^{\dagger}
a_{\overline{\sigma}}^{\dagger},
\nn \\  
&&
S^z=\frac{1}{4}\sum^s_{\sigma=-s}\left(2a^\dagger_{\s}a_{\s}
+(-1)^{2s} I\right), \label{idr} 
\eea   
satisfying (\ref{su2}) where $\overline{\sigma}=-\sigma$. 
In each case the vacuum state $\left|0\right>$ is a lowest weight state
of weight $-\S=(-1)^{2s} \frac14 (2s+1)$. 
In the fermionic case (where $s$ is a half-odd integer) these
representations are finite-dimensional  and $\S=\frac12,1,\frac32,\dots$. 
In the bosonic case (where $s$ is an integer) the representations are
infinite-dimensional and 
$ \S=-\frac14,-\frac34,-\frac54,\dots$. For the bosonic case, we can also consider
lowest weight states of the form 
\beq 
a^\dagger_{\sigma_1}\left|0\right>, \qquad
a^\dagger_{\sigma_1} a^\dagger_{\sigma_2} \left|0\right> , \qquad
a^\dagger_{\sigma_1} a^\dagger_{\sigma_2} a^\dagger_{\sigma_3}
\left|0\right>, \qquad {\rm etc},
\label{bosons}  
\eeq 
provided $\sigma_i\neq \overline{\sigma_j}$ for all $i,\,j$. 
When there are
$z$ bosons in a lowest weight state of the form (\ref{bosons}) 
the lowest weight is given by 
$-\S=\frac14 (2z+(-1)^{2s}(2s+1))$ 
giving rise to the sequence 
$\S=-\frac12 z-\frac14,-\frac12 z-\frac34,-\frac12 z-\frac54,\dots$.  
Under these representations (\ref{ham}) describes  pairing Hamiltonians for
spin $s$ particles, and the integer $m$ in (\ref{bcsbae}) denotes the number
of Cooper pairs in the corresponding eigenstate.
The usual reduced BCS model for spin $\frac12$ fermions corresponds to the case
$s=\frac12$ giving $\S=\frac12$ \cite{lzmg,yba}.

Following \cite{yba} we consider the large $g$ limit. Let $r\leq m$ denote
the number of roots $E_j$ which diverge as $g\rightarrow \infty$.  
For these roots we have to lowest order in $g^{-1}$  
\beq
\frac{2}{g}+ \frac{n}{E_j}  
=\sum_{\ell\neq j}^r \frac{2}{E_j-E_\ell}.
\label{limbae}  \eeq
where $j=1,\ldots,r$ and
\beq n=2\S\L+2r-2m.  \label{rel} 
\eeq  
The remaining roots $E_j$, $j=r+1,\ldots,m$, are small in comparison with $g$.
Comparison between      
(\ref{bose}), (\ref{bose2}) and (\ref{limbae}) shows we may identify 
$E_j=k_{2j-1}^2=k_{2j}^2$ for $N$ even and 
$E_j=k_{2j}^2=k_{2j+1}^2$ for $N$ odd, with 
\beq
\quad \frac{2}{g}=\frac{L}{2c}, \quad r=M,
\eeq
where $n=-\frac{1}{2}$ for $N$ even, and $n=-\frac{3}{2}$ for $N$ odd. 
The fact that $\L,\,m$ are positive integers, $r$ is a non-negative
integer and $r\leq m$ imposes severe restrictions on the allowed
solutions of (\ref{rel}). 
For $n=-\frac12$ the only solution is $\L=1,\,m=M$ and $\S=-\frac14$ ($s=0,\,z=0$). 
For $n=-\frac32$ we can have $\L=3,\,m=M$ and $\S=-\frac14$ ($s=0,\,z=0$).  
In all cases $s$ is an
integer so the mapping is to a bosonic pairing model. Moreover, since
$g$ is positive, and the bosonic representations (\ref{idr}) are
non-unitary (specifically $(S^+)^\dagger =-S^-$) these systems
describe {\it repulsive} pairing interactions. 
In the strong coupling limit, all energy levels may collapse into one
multiply-degenerate level for $N$ even and into three energy levels for $N$ odd.
In comparison with large pairing scattering energies, the level spacing is negligible. 
Therefore the multiply-degenerate levels are a reasonable expectation 
in the strong coupling limit.

Multiplying each of (\ref{limbae}) by $E_j$ and taking the sum gives 
\bea
E(r)=\sum_{j=1}^r E_j&=&gr(r-n-1) \nn \\ &=&-gr(2\S\L+r-2m+1).
\eea 
The energy  
function $E(r)$ has zeroes at $0$ and $2m-2\S\L-1$ and attains the maximum
value at $m-\S\L-\frac12$. 
Note that for bosonic pairing models 
$-\S\geq \frac14 +\frac{z}{4}$ ($z=0$ in this case), 
which imposes the lower bound $m-\frac14$ for the value at which the
maximum occurs. The energy gaps between successive levels are found to be
\bea 
\Delta(r)&=&E(r)-E(r+1)  \nn \\
&=& 2g(\S\L+r+1-m), 
\label{gaps} 
\eea
and in particular 
\bea 
\Delta({m-1})&=&2g\S\L,    \nn \\
\Delta({m-2})&=&2g(\S\L-1),    \nn \\
\Delta({m-3})&=&2g(\S\L-2).     
\eea 
%
We see that $\Delta(r) < 0$ if $r \in \{0,1,\ldots,m-1 \}$ and 
$\S\le -\frac{z}{2}-\frac14$.
Thus $E(r)$ takes its maximal value when $r=m=M$.
%
%
>From Eqs.~(\ref{bose}) and (\ref{bose2}), the
ground state energy per particle of the weakly interacting Bose gas follows as 
\begin{equation}
\frac{{\cal E}_0}{N}=\sum_{i=1}^{N}\frac{k_i^2}{N}=\frac{2 E(M)}{N}=\frac{c(N-1)}{L}.
\end{equation}
This agrees with Eq.~(\ref{enrgy1}) and the results of \cite{LL,TAK,Wadati} among others.
We see that it also coincides with the highest energy per particle state
of the strongly coupled BCS boson pairing model (\ref{ham}).

In conclusion we have considered the asymptotic solutions to the Bethe ansatz equations for the 
the weakly interacting 1D Bose gas. 
We have established that the ground state 
maps to the highest energy 
state of a strongly-coupled repulsive bosonic pairing model.  
However, the precise link between the integrable boson model with weakly repulsive 
delta interaction and the standard BCS fermionic model is quite subtle and  
deserves further investigation.

\section*{Acknowledgment}

This work has been supported by the Australian Research Council.
MTB thanks the organisers of The International Conference on the Statistical Physics of Quantum Systems for the opportunity to present this work. 
We thank Huan-Qiang Zhou  for helpful discussions
and the referee for some helpful remarks.

\end{document}